\documentclass[aps,reprint,amsmath,amssymb,superscriptaddress]{revtex4-2}

\usepackage{graphicx}
\usepackage{hyperref}
\usepackage{bm}
\usepackage{braket}

\begin{document}

\title{Self-organized PT-symmetry of exciton-polariton condensate in a double-well potential}

\author{P.A. Kalozoumis}
\affiliation{Materials Science Department, School of Natural Sciences, University of Patras, GR-26504 Patras, Greece}
\affiliation{Hellenic American University, 436 Amherst st, Nashua, NH 0306 USA}
\affiliation{Institute of Electronic Structure and Laser, FORTH, GR-70013 Heraklion, Crete, Greece}

\author{D. Petrosyan}
\affiliation{Institute of Electronic Structure and Laser, FORTH, GR-70013 Heraklion, Crete, Greece}
\affiliation{A. Alikhanyan National Science Laboratory (YerPhI), 0036 Yerevan, Armenia}

\date{\today}

\begin{abstract}
We investigate the dynamics and stationary states of a semiconductor exciton-polariton condensate 
in a double well potential. We find that upon the population build up of the polaritons by above-threshold 
laser pumping, coherence relaxation due to the phase fluctuations of the polaritons drives the system into 
a stable fixed point corresponding to a self-organized PT-symmetric phase.
\end{abstract}

\maketitle

\section{Introduction}
\label{sec:intro} 

One of the prominent research directions in semiconductor optics is the study of exciton-polariton condensation 
in microcavities. Exciton-polaritons are hybrid quasi-particles of strongly coupled quantum well excitons
and cavity photons~\cite{DengRevModPhys2010,CarusottoRevModPhys2010} which retain the properties of both matter and light. 
The excitonic part mediates effective interactions between the polaritons, 
giving rise to interesting nonlinear properties, whereas the small effective mass of the photonic component enables 
Bose-Einstein condensation even at ambient temperatures~\cite{LagoudakisNatPhys2008,YamaguchiPRL2013,SchneiderNature2013},
in contrast to their ultracold atomic counterparts~\cite{GoblotPRL2016}. 
The short lifetime of the polariton condensate renders it an open system
that requires continuous replenishing from the excitonic reservoir via external pumping. 
After their experimental realization~\cite{KasprzakNature2006,BaliliScience2007}, the polariton condensates have 
been shown to be ideal system for studies of many effects at the interface of non-equilibrium physics and nonlinear dynamics. 

The intrinsic nonlinear dynamics of polariton systems lead to a variety of effects, such as 
the appearance of a Mach-Cherenkov cone in a supersonic flow~\cite{AmoNatPhys2009}, 
the formation of quantized vortices~\cite{DominiciNatComm2018}, and dark solitons~\cite{GonzalezPhysLettA2017}. 
Moreover, the polariton condensates can be engineered with high precision by the external laser fields 
\cite{BaliliScience2007,Ohadi2017, Ohadi2018, Orfanakis2021}. Finally, such systems are promising candidates 
for various applications in photonic devices, such as switches, gates and transistors~\cite{LagoudakisNatPhys2008}, 
as well as for quantum simulators of interacting spin models~\cite{BerloffNatMat2017}.

The ``open'' nature of the system, featuring gain and loss, leads to interesting implications 
when the dissipative dynamics become pseudo-Hermitian. This is the case in parity-time (PT) symmetric setups, 
where dissipation losses are exactly balanced by the pumping gain.
Systems with PT-symmetry has been a flourishing and broad research field, 
extending from quantum mechanics~\cite{Bender2002} and field theory~\cite{Bender2004} 
to optics~\cite{Ganainy2007} and acoustics~\cite{Fleury2015}.  

The interplay between the inherent losses and the laser pumping in such a way as to preserve the PT symmetry 
of the system provides an effective framework where a polariton system can exhibit coherent, Hermitian-like dynamics 
for relatively long times. Recent works have shown promising results, such as permanent Rabi 
oscillations~\cite{ChestnovSciRep2016}, multistability and condensation below threshold~\cite{LienPRB2015}, 
exceptional points in polaritonic cavities below lasing threshold~\cite{KhurginOPTICA2020}, and coherent oscillations 
of a two species polariton mixture in a double well~\cite{KalozoumisEPL2020}. 
The latter has been shown to be able to simulate the dynamics of a pair of spin-1/2 particles (qubits) in the presence 
of exchange interaction. Yet, polariton structures in the framework of PT symmetry have not been extensively 
studied yet, and more efforts are required to understand the rich landscape of phenomena which emerge from this framework. 

In this work we study the dynamics of an exciton-polariton condensate in a double well potential, 
in the presence of time-varying exciton populations and phase fluctuations. 
We consider the coupled-mode equations for the polaritons supplemented by the rate equations for the laser-pumped exciton
reservoirs, and derive analytically the steady state solutions for the exciton and polariton populations as well as their 
coherence. We find that, when the total pumping rate is above threshold, the system automatically attains the PT 
symmetric state, independently of the pumping rates of the individual sites. Employing numerical simulations for several 
different pumping rates and initial conditions, we verify our analytical findings. We also study the stability and 
robustness of our results in the presence of phase noise caused by the unavoidable phase fluctuations of the polaritons. 


\section{The exciton-polariton system}\label{II}

\begin{figure}[t]
\includegraphics[width=1.0\columnwidth]{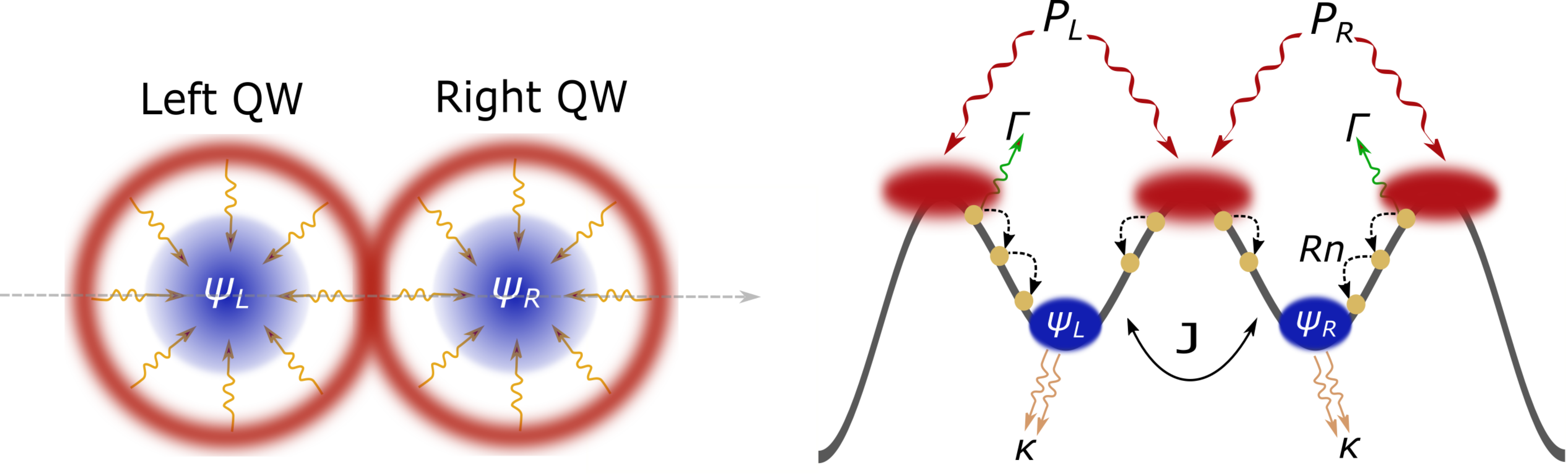}
\caption{Schematic top view (left panel) and side view (right panel) of a polariton system in a double quantum well. 
  Spatially shaped pumping lasers populate with rates $P_L$ and $P_R$ the reservoir excitons $n_{L,R}$,  which 
  decay via recombination with rates $\Gamma$ and energy-relax and scatter into the polariton condensate with rate $R$.
  The pumping lasers also create the confining potentials for the polaritons $\psi_L$ and $\psi_R$, which decay 
  with rates $\kappa$, are continuously replenished by reservoir excitons with rates $Rn_{L,R}$, while interacting 
  with each other via the Josephson coupling $J$.} 
\label{fig1}
\end{figure}

The system under consideration is schematically illustrated in Fig.~\ref{fig1}. 
One or more layers of semiconductor quantum wells are placed inside the semiconductor microcavity near 
the antinode of the resonant cavity field mode. Spatially shaped pumping lasers replenish continuously the exciton 
reservoirs and simultaneously create confining potentials for the polariton condensate. 
Assuming a tight-binding double-well potential, the exciton-polariton system can be described by the following set 
of equations for the polariton condensate wavefunctions $\psi_{L}$ and $\psi_{R}$ in the left ($L$) and right ($R$) 
wells \cite{WoutersPRL2007}:
\begin{subequations}
\label{full_model}
\begin{eqnarray}
i \partial_t \psi_{L} &=& \left[ \epsilon_L  + \eta |\psi_{L}|^{2} \right] \psi_L 
+ \frac{i}{2} \left[ R n_{L} - \kappa \right] \psi_L 
- J \; \psi_{R} , \qquad \label{site_one}  \\
 i \partial_t \psi_{R} &=& \left[ \epsilon_R + \eta |\psi_{R}|^{2} \right] \psi_R 
+ \frac{i}{2} \left[ R n_{R} - \kappa \right] \psi_R 
- J^{*}\psi_{L} ,  \label{site_two} 
\end{eqnarray}
\end{subequations}
where $\epsilon_{L,R}$ are the single-particle energies, $\eta$ is the nonlinear interaction strength,
$\kappa$ is the decay rate of the polaritons due to the exciton recombination and cavity photon losses 
(assumed the same for both wells), and $J$ is the Josephson (tunnel) coupling between the wells.
The polariton equations are supplemented by the equations for the populations $n_{L,R}$ of the reservoir excitons, 
\begin{subequations}
\label{resLR}
\begin{eqnarray}
 \partial_t n_{L} &=& P_{L} - \Gamma n_{L} -R n_{L} |\psi_{L}|^2 , 
\label{res_one} \\
 \partial_t n_{R} &=& P_{R} - \Gamma n_{R} -R n_{R} |\psi_{R}|^2 ,
 \label{res_two} 
\end{eqnarray}
\end{subequations}
which are created by laser pumping with rates $P_{L,R}$, decay with rate $\Gamma$, and scatter into the polariton 
condensate with rate $R$.    

In Appendix~\ref{sec:app} we briefly outline the PT-symmetry conditions for a condensate in a double 
well potential. Neglecting for the moment the non-linearity $\eta$, the PT-symmetry condition is satisfied 
when $\epsilon_{L,R} = \epsilon \, (=0)$ and the gain in one well exactly compensates the losses in the other,
\begin{equation}
\gamma_L \equiv \frac{1}{2}[R n_{L} - \kappa] = - \frac{1}{2}[R n_{R} - \kappa] \equiv -\gamma_R , \label{gammadefs}
\end{equation}
as per Eqs.~(\ref{site_one}) and ~(\ref{site_two}), which leads to  
$n_{L} + n_R = \frac{2\kappa}{R}$.
The threshold pumping at which the polariton condensate starts to form can be obtained from 
the condition that the sum of the gain and loss in both wells is non-negative, $\gamma_L + \gamma_R \geq 0$.
With Eq.~(\ref{gammadefs}), this condition is equivalent to 
\begin{equation}
\label{Thexcitons}
n_{L} + n_R \geq \frac{2\kappa}{R}.
\end{equation}
Note that exactly at the threshold, this is the same condition as for the PT-symmetry.
If we consider the stationary regime for the reservoir excitons, $\partial_t n_{L,R}=0$,  
we find from Eqs.~(\ref{resLR}) the steady-state values 
\begin{equation}
n_{L,R} = \frac{P_{L,R}}{\Gamma+R|\psi_{L,R}|^2} . \label{excitons_stat_1}
\end{equation}

Exactly at the threshold for condensate formation, the values of the polariton populations
in both wells, $|\psi_{L,R}|^2$, are marginally equal to zero and we have $n_{L,R} \simeq P_{L,R}/\Gamma$. 
Substituting these values into Eq.~(\ref{Thexcitons}), we find the threshold pumping condition
\begin{equation}
\label{threshold_pumping} P_L+P_R \geq \frac{2\kappa \Gamma}{R} ,
\end{equation}
above which the condensate begins to form, while 
for $P_L+P_R<2\kappa \Gamma/R$ the condensate decays to zero.

\begin{figure*}[t]
\includegraphics[width=0.9\textwidth]{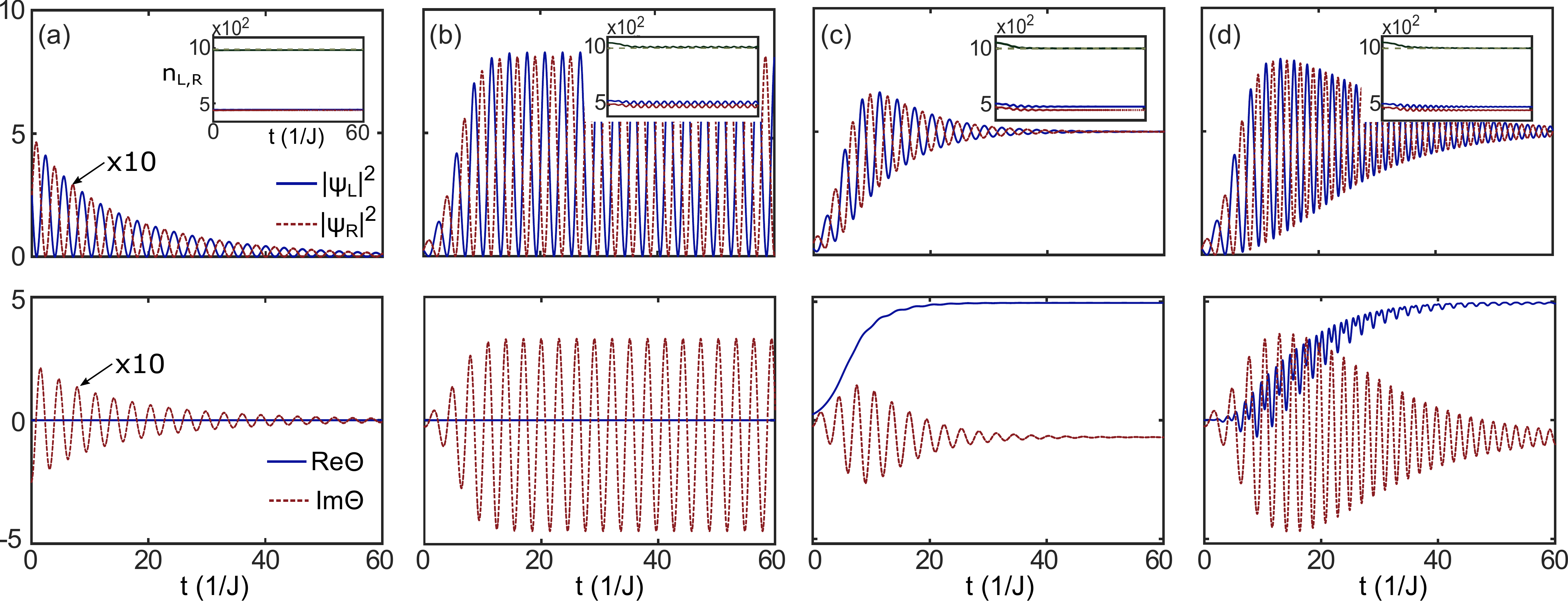}
\caption{Dynamics of the polariton populations $|\psi_{L,R}|^2$ (upper panels), the reservoir excitons $n_{L,R}$ (insets), 
and the coherence $\Theta$ (lower panels), as obtained from the numerical solution of Eqs.~(\ref{full_model}-\ref{resLR})
for the parameters $\epsilon =0$, $\kappa=10J$, $\Gamma=2J$, $R=0.02J$ (time is in units of $J^{-1}$), 
corresponding to the threshold values of pumping $P_L+P_R = 2\kappa \Gamma/R = 2000J$ and 
the steady-state exciton populations $n_L+n_R = 2\kappa /R = 1000$ (green dashed lines in the insets).
The initial exciton populations are always taken as $n_{L,R} = P_{L,R}/\Gamma$, while 
the initial polariton amplitudes $\psi_{L,R}$ are seeded with small (complex) values.    
(a)~Linear case ($\eta=0$) with the pumping rates $P_{L}=1000J$ and $P_R=990J$ slightly below threshold, 
leading to decay of the initial polariton populations $|\psi_{L}(t)|^2$ and $|\psi_{R}(t)|$ and 
steady-state exciton population $n_{L}+n_{R} < 2\kappa/R$. 
(b)~Same as in (a), but for stronger pumping rates $P_L=1080$ and $P_R=1020$ above threshold and
initial conditions $\textrm{Re}\Theta(0) = 0$, leading to the initial build up of 
the polariton populations and their continuous Rabi-like oscillations, while $\textrm{Re}\Theta(t) = 0 \; \forall \; t$,
and the exciton population $n_{L}+n_{R}$ oscillating slightly above the threshold value $2\kappa/R = 1000$.
(c)~Same as in (b) but for the initial conditions $\textrm{Re}\Theta(0) \neq 0$,
leading to a steady-state of the system. 
(d)~Same as in (b) with the initial conditions $\textrm{Re}\Theta(0) = 0$ but in the presence of nonlinearity $\eta = 0.3J$ 
that couples $\textrm{Re}\Theta$ and $\textrm{Im}\Theta$, leading to a steady-state of the system. } 
\label{fig2}
\end{figure*}

In the upper panels of Fig.~\ref{fig2} we show the polariton populations $|\psi_{L,R}|^2$ 
for different pumping rates and initial conditions, with and without non-linear interaction,
as obtained from the numerical solution of Eqs.~(\ref{full_model}-\ref{resLR}). 
The insets show the evolution of the exciton populations $n_{L}$ and $n_{R}$ and their sum $n_{L}+n_{R}$. 
For pumping below threshold, we observe a decay of the initial (seed) polariton populations with rate
$\gamma_L+ \gamma_R \simeq \frac{R}{2\Gamma} (P_L + P_R) -\kappa < 0$, 
accompanied by Rabi-like oscillations, while the exciton populations settle to $n_{L,R}=P_{L,R}/\Gamma$. 
For pumping above threshold, the polariton populations grow until reaching certain values $|\psi_{L,R}|^2$ at
which $n_{L} + n_R \simeq \frac{2\kappa}{R}$, while the Rabi-like oscillations persist or are eventually damped,
depending on the initial conditions or presence of non-linear interaction, as discussed below. 
Remarkably, the polariton and exciton populations increase and decrease, respectively, 
reaching the same stationary values which satisfy the PT-symmetry conditions, 
independently of the pumping rates, as long as pumping is retained above threshold. 

\section{Equivalence of the PT-symmetry and steady state conditions}
\label{sec:PTSS}

To understand the dynamics of the system, it is convenient to express Eqs.~(\ref{full_model})
in terms of the polariton populations $|\psi_{L,R}|^2$ and coherence $\Theta \equiv \psi_L \psi_{R}^*$ as 
\begin{subequations}
\label{population_coherence}
\begin{eqnarray}
\partial_t |\psi_L|^2 &=& 2 \gamma_{L} |\psi_{L}|^2 + 2J \textrm{Im} \Theta
\label{population_coherence1} , \\
\partial_t |\psi_R|^2 &=& 2 \gamma_{R} |\psi_{R}|^2 - 2J \textrm{Im} \Theta
\label{population_coherence2} , \\
\partial_t \Theta &=& -i \left[ \epsilon_L - \epsilon_R + \eta ( |\psi_{L}|^2 -|\psi_{R}|^2 ) \right] \Theta
\nonumber \\ & & 
+ (\gamma_L+ \gamma_R) \Theta - i J \left(|\psi_{L}|^2  - |\psi_{R}|^2  \right) \label{population_coherence3} .
\end{eqnarray}
\end{subequations}
Note that below threshold, $(\gamma_L+ \gamma_R) < 0$, both the polariton populations and their coherence decay to zero,
as already mentioned above.   


Let us assume $\epsilon_{L,R} = 0$ and consider first the case of vanishing nonlinearity $\eta=0$.
Equation~(\ref{population_coherence3}) indicates that the coherence decays only if its real part is nonzero.
In turn, the solution for the real part of the coherence is  
\begin{equation}
\label{real_coh_sol} \textrm{Re} \Theta (t) =  \textrm{Re} \Theta (0) 
\; e^{ \int_0^t (\gamma_L + \gamma_R) dt'} .
\end{equation}
Hence, if initially $\textrm{Re} \Theta(0) = 0$, it will remain so at later times, $\textrm{Re}\Theta(t) =0 \; \forall \; t >0$. 
Then the dynamics of the system, if pumped above threshold, will exhibit continuous Rabi-like oscillations 
with frequency $J$, while no steady state will be attained, as in Fig.~\ref{fig2}(b).

In practice, however, even if initially we have $\textrm{Re} \Theta(0) = 0$ 
[e.g., either $\psi_{L}(0) = 0$ or $\psi_{R}(0) = 0$], 
the unavoidable phase fluctuations of the polaritons will eventually lead to the appearance of finite
$\textrm{Re} \Theta(t) \neq 0$, which in turn will result in the decay of coherence 
and drive the system to the steady state. Equivalently, if we have initially 
$\textrm{Re} \Theta(0) \neq 0$, the system can still exhibit initially Rabi-like oscillations,
but then it will eventually attain the steady state, as in Fig.~\ref{fig2}(c).
Finally, as seen from Eq.~(\ref{population_coherence3}) the nonlinear interaction couples the real and imaginary 
parts of the coherence $\Theta$ with the rate $\eta (|\psi_{L}|^2 -|\psi_{R}|^2)$. 
Hence, in the presence of nonlinearity $\eta \neq 0$, we expect 
the eventual decay of the coherence with the system attaining the steady state, for any initial conditions
and independent on the phase fluctuations, as in Fig.~\ref{fig2}(d).

Setting the time derivative in the left-hand side of the Eq.~(\ref{population_coherence3}) equal to zero, 
we find the steady state is reached when 
\begin{equation}
\label{stationary_cond_1} R[n_{L}+n_{R}]-2 \kappa = 0 , \quad \mathrm{and} \quad |\psi_{L}|^2 = |\psi_{R}|^2. 
\end{equation}
Remarkably, the first equation corresponds exactly to the PT-symmetry condition $n_{L} + n_R = \frac{2\kappa}{R}$
discussed above. Moreover, this condition is satisfied even in the presence of nonlinear interaction $\eta \neq 0$, 
because the equal polariton populations as per the second equation lead to exactly the same energy shifts $\eta |\psi_{L,R}|^2$ 
of the polaritons in both wells. In other words, for any initial conditions, and provided the total pumping is 
above threshold as per Eq.~(\ref{threshold_pumping}) but otherwise arbitrary $P_L$ and $P_R$, the system 
attains a stable fixed point corresponding to the PT-symmetric state. Even when no steady state exists 
or is yet reached, the PT condition in Eq.~(\ref{stationary_cond_1}) is approximately satisfied, 
as seen in the insets of Fig.~\ref{fig2}.

\begin{figure*}[t]
\centering \includegraphics[width=0.8\textwidth]{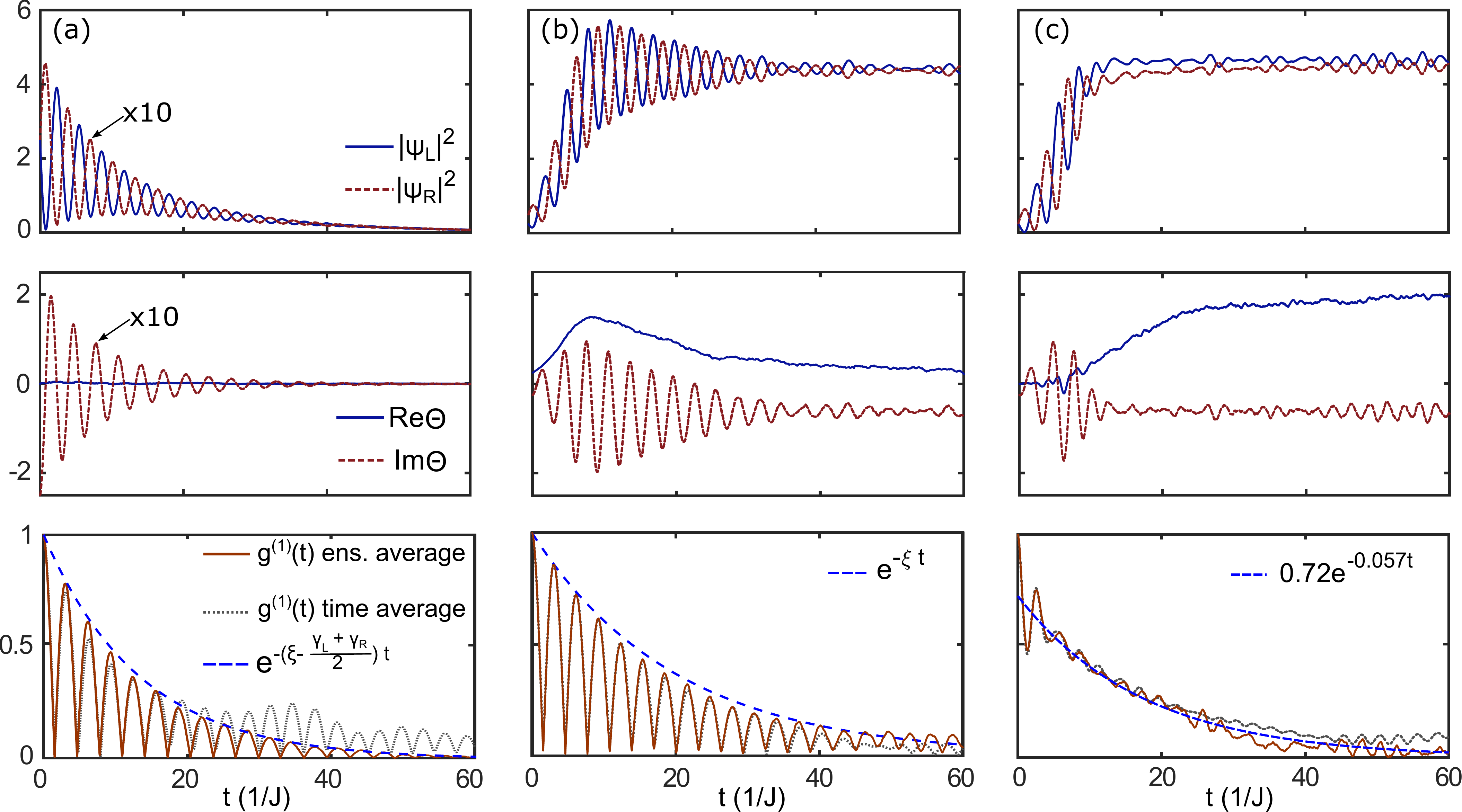}
\caption{Dynamics of the polariton populations $|\psi_{L,R}|^2$ (upper panels), the coherence $\Theta$ (middle panels),
and the left-well polariton field correlation function $g^{(1)}(t)$ (lower panels), for the same parameters as in Fig.~\ref{fig2}
with the addition of phase fluctuations causing decoherence with rate $\xi=0.05J$, as obtained from the the ensemble
averaged solution of Eqs.~(\ref{full_model}-\ref{resLR}).   
(a)~Linear case ($\eta=0$) with the below threshold pumping $P_{L}=1000J$ and $P_R=990J$ leading to the 
exponential decay of the correlation function $g^{(1)}(t) \propto e^{-(\xi- |\gamma_L+\gamma_R|/2)t}$.
(b)~Same as in (a), but for stronger pumping above threshold $P_L=1080$ and $P_R=1020$. 
Now, independently of the initial value of $\textrm{Re} \Theta(0)$, the phase fluctuations
cause exponential decay of $g^{(1)}(t) \propto e^{-\xi t}$ while the system approaches the steady state.
(c)~Same as in (b) but in the presence of nonlinearity $\eta=0.3J$, causing accelerated decay of
$g^{(1)}(t) \propto e^{-0.057t}$ and faster approach of the system to the steady state. 
In the lower panels, we also show the correlation functions $g^{(1)}(t)$ obtained from the long-time
average of the system dynamics (the oscillating tail of $g^{(1)}$ is due to the finite length of the time series).}
\label{fig3}
\end{figure*}

Combining Eqs.~(\ref{excitons_stat_1}) and (\ref{stationary_cond_1}), we find that steady state 
polariton populations are
\begin{equation}
\label{stationary_polaritons} |\psi_{L}|^2=|\psi_{R}|^2=\frac{P_L+P_R}{2\kappa}-\frac{\Gamma}{R},
\end{equation}
while the exciton populations are 
\begin{equation}
\label{excitons_stat_2} n_{L,R}=\frac{2\kappa P_{L,R}}{R(P_L+P_R)}.
\end{equation}
Using these stationary values for $n_{L,R}$ and $|\psi_{L,R}|^2$ in 
Eqs.~(\ref{population_coherence1}) and (\ref{population_coherence2}) in the steady state we obtain
\begin{equation}
\label{imag_coherence_stat3}
\textrm{Im} (\Theta) = \frac{\kappa \Gamma ( P_{L}-P_R) }{2RJ(P_L+P_R)}-\frac{P_{L}-P_R}{4J} ,
\end{equation}
and from $|\psi_L \psi_R^*|^2= [ \textrm{Re}(\psi_{L} \psi_{R}^*) ]^2+ [ \textrm{Im}( \psi_{L} \psi_{R}^*) ]^2$
we obtain 
\begin{equation}
\label{real_coherence_stat} 
\textrm{Re}(\Theta)=\mathcal{D}\sqrt{4J^2(P_L+P_R)^2-\kappa^2 (P_L-P_R)^2}
\end{equation}
where 
\[
\mathcal{D}=\frac{P_L+P_R-2\kappa \Gamma/R}{4J\gamma(P_L+P_R)} .
\]
These results are verified by the numerical simulations illustrated in Fig.~\ref{fig2} 
and they equally hold for any value the nonlinearity strength $\eta$.

\section{Phase fluctuations}
\label{sec:PF}

As mentioned above, the coherence of the polariton condensate will decay due to the phase fluctuations that 
are always present in realistic quantum systems. We therefore incorporate the phase noise in our numerical 
calculations and investigate how it modifies the dynamics of the polaritons and the coherence.
We model the phase fluctuations as the standard Wiener process for stochastic differential equations. 
Thus, the single-particle energies $\epsilon_{L,R}$ in Eqs.~(\ref{full_model}) become Gaussian 
stochastic variables with the mean $\braket{\epsilon_{L,R}} = 0$ and variance $\sigma^2 = 2 \xi/\delta t$, 
where $\xi$ is the decoherence rate and $\delta t$ is the time step for picking a new random energy.

In Fig.~\ref{fig3} we show the results of our numerical simulations as obtained upon the ensemble average 
over $N=1000$ independent realizations of the system dynamics. We compute the first-order correlation
functions $g^{(1)}(t)$, which quantify the coherence for the polaritonic fields, via  
\begin{equation}
\label{acf_ens}
g^{(1)}(t)= \frac{\langle \psi(t_0)\psi(t)\rangle}{\sqrt{\langle |\psi(t_0)|^2 \rangle \langle |\psi(t)|^2 \rangle}} ,
\end{equation}
where $\psi = \psi_L$ or $\psi_R$, and $\braket{\ldots}$ denotes the ensemble average. 

Below the pumping threshold, the polariton fields decay with rate $\gamma_L + \gamma_R < 0$,
but the phase fluctuations with rate $\xi$ causes even faster decay of coherence, 
$g^{(1)}(t) \propto e^{-(\xi-|\gamma_L+\gamma_R|/2)t}$, as seen in Fig.~\ref{fig3}(a).
For pumping above threshold, the phase noise causes exponential decay of the correlation function $g^{(1)}(t) \propto e^{-\xi t}$, 
while the system approaches the steady state independently of the initial value of coherence $\textrm{Re} \Theta(0)$, 
as seen in Fig.~\ref{fig3}(b). Including also the nonlinear interaction $\eta \neq 0$ further accelerates the decay 
of the correlation function and the system approaches the steady state even faster. 

We finally note that for an ergodic system the ensemble-averaged and time-averaged correlation functions are equivalent. 
To verify whether our polariton system is ergodic, we also compute the field correlation function
\begin{equation}
\label{acf_time}
g^{(1)}(t)=\frac{ \int_{t_\textrm{i}}^{t_{\textrm{f}}} d \tau \psi (\tau) \psi^{*}(\tau+t)}
{\sqrt{\int_{t_\textrm{i}}^{t_{\textrm{f}}} d \tau |\psi(\tau)|^2 \int_{t_\textrm{i}}^{t_{\textrm{f}}} d \tau |\psi(\tau+t)|^2}} 
\end{equation}
resulting from a single, long-time trajectory with $t_{\textrm{f}} -t_{\textrm{i}} = 3000/J$.  
As seen in Fig.~\ref{fig3} (lower panels), the computed ensemble-averaged and time-averaged correlation functions 
coincide to a very approximation, attesting to the ergodicity of our system.

\section{Conclusions}
\label{sec:conc}

To summarize, we have studied an exciton-polariton system in a double-well potential, 
taking into account the dynamics of the reservoir excitons and the polaritons.
We have found that for pumping of the excitons above the total threshold value 
for the formation of the polariton condensate, the exciton populations attain the 
values that satisfy the PT-symmetry condition for the polariton condensate, independent 
of the pumping rates of the individual wells. Employing the population-coherence equations,
we interpreted the corresponding dynamics and revealed the stable fixed point, or the steady state, that the system approaches. 
To make our analysis experimentally relevant, we have taken also into account the phase fluctuations
present in any realistic system, and computed the first-order correlation functions for the
polariton fields, which revealed the coherence decay with the corresponding rate. 

We note that our results apply to moderate non-linear interaction strength and small 
differences in pumping rates of the two wells. For large difference in the pumping rates,
the strong non-linear energy shift of the polariton condensate energy may lead to
self-trapping and break-up of the PT symmetry \cite{Sukhorukov2010}

\section{Acknowledgments}
We thank P.G. Savvidis, H. Ohadi, and A.F. Tzortzakakis for fruitful discussions. 
This work was co-financed by Greece (General Secretariat for Research and Technology), 
and the European Union (European Regional Development Fund), in the framework of the bilateral 
Greek-Russian Science and Technology collaboration on Quantum Technologies (POLISIMULATOR project.)

\appendix
\section{Polariton condensate in a PT-symmetric double well}
\label{sec:app}

Consider a polariton condensate in a double well potential described by the coupled-mode equations
\begin{subequations}
\label{eqs:pisLR}
\begin{eqnarray}
\label{site_one_plus}
 i \partial_t \psi_{L} &=& (\epsilon_L + i \gamma_L) \psi_{L} + \eta |\psi_{L}|^{2}\psi_{L} - J \; \psi_{R}  , \\
\label{site_one_minus}
 i \partial_t \psi_{R} &=& (\epsilon_R + i \gamma_R) \psi_{R} + \eta |\psi_{R}|^{2}\psi_{R} - J^* \! \psi_{L} ,
\end{eqnarray}
\end{subequations}
where $\epsilon_{L,R}$ are the single-particle energies, 
$\gamma_{L,R}$ are the incoherent loss ($\gamma <0$) or gain ($\gamma >0$) rates at each well, 
$\eta$ is the nonlinear interaction strength, and $J$ is the Josephson coupling between the wells. 

\begin{figure}[t]
\centering \includegraphics[width=0.6\columnwidth]{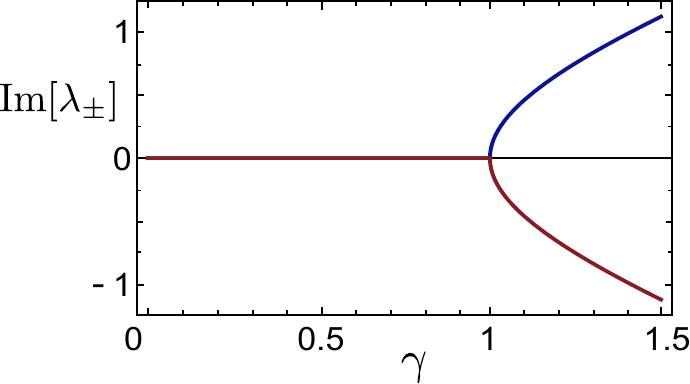}
\caption{Imaginary part of the eigenvalues $\lambda_{\pm}$ in Eq.~(\ref{eigenvalues}). 
For $\gamma < J (=1)$ we have a PT symmetric phase with real eigenvalues.
Above the bifurcation point at $\gamma=J$, the system enters the PT broken phase
with imaginary eigenvalues. }
\label{fig_app}
\end{figure}

If we set $\epsilon_{L,R} =0$, assume negligibly weak nonlinearity, $\eta |\psi|^{2} \ll J$, and 
set $\gamma_{L} = - \gamma_{R}= \gamma$ so that the loss at the right well is exactly compensated by the gain at the left well, 
we obtain a PT-symmetric Hamiltonian matrix corresponding to Eqs.~(\ref{eqs:pisLR}) \cite{KalozoumisEPL2020}:   
\begin{equation}
\label{hamiltonian} \mathcal{H} =  \begin{pmatrix}
 i \gamma & - J \\  - J^* & - i\gamma 
\end{pmatrix}.
\end{equation}
Its eigenvalues and the corresponding eigenvectors are given by
\begin{equation}
\label{eigenvalues}
 \lambda_{\pm} = \pm \sqrt{|J|^2-\gamma^2}
\end{equation}
and
\begin{equation}
\label{eigenstateS}
|\pm \rangle = \left[ \left( \sqrt{|J|^2-\gamma^2} \pm i \gamma \right) | L \rangle \mp J^*  | R \rangle \right]/N_{\pm}
\end{equation}
with $N_{\pm}$ the normalization factors.
For $\gamma < |J|$, the eigenvalue spectrum is real and the dynamics is Hermitian-like. 
For $\gamma > |J|$, the eigenvalues become imaginary and the system enters the PT-broken phase.
The case $|J|=\gamma$ corresponds to the exceptional point of the system where the eigenvalues become degenerate 
and the eigenstates coalesce.
Figure~\ref{fig_app} illustrates the dependence of imaginary part of the eigenvalues on the loss/gain parameter $\gamma$.

\end{document}